# Restored strange metal phase through suppression of charge density waves in underdoped YBa$_2$Cu$_3$O$_{7-\delta}$

**Authors:** Eric Wahlberg[1], Riccardo Arpaia[1,2,*], Götz Seibold[3], Matteo Rossi[2,†], Roberto Fumagalli[2], Edoardo Trabaldo[1], Nicholas B. Brookes[4], Lucio Braicovich[2,4], Sergio Caprara[5,6], Ulf Gran[7], Giacomo Ghiringhelli[2,8], Thilo Bauch[1], Floriana Lombardi[1,*]

**Affiliations:**

[1] Quantum Device Physics Laboratory, Department of Microtechnology and Nanoscience, Chalmers University of Technology, SE-41296, Göteborg, Sweden

[2] Dipartimento di Fisica, Politecnico di Milano, I-20133 Milano, Italy

[3] Institut für Physik, BTU Cottbus-Senftenberg - PBox 101344, D-03013, Cottbus, Germany

[4] ESRF, The European Synchrotron, F-38043 Grenoble, France

[5] Dipartimento di Fisica, Università di Roma "La Sapienza", P.$^{le}$ Aldo Moro 5, I-00185 Roma, Italy

[6] CNR-ISC, via dei Taurini 19, I-00185 Roma, Italy

[7] Division of Subatomic, High-Energy and Plasma Physics, Chalmers University of Technology, SE-41296, Göteborg, Sweden

[8] CNR-SPIN, Dipartimento di Fisica, Politecnico di Milano, I-20133 Milano, Italy

[†] Present address: Stanford Institute for Materials and Energy Sciences, SLAC National Accelerator Laboratory, Menlo Park, CA-94025, USA

[*] Corresponding author. E-mail: floriana.lombardi@chalmers.se (F.L.); riccardo.arpaia@chalmers.se (R.A.)

**Abstract:** The normal state of optimally doped cuprates is dominated by the "strange metal" phase that shows a linear temperature ($T$) dependence of the resistivity persisting down to the lowest $T$. For underdoped cuprates this behavior is lost below the pseudogap temperature $T^*$, where Charge Density Waves (CDW) together with other intertwined local orders characterize the ground state. Here we show that the $T$-linear resistivity of highly strained, ultrathin and underdoped YBa$_2$Cu$_3$O$_{7-\delta}$ films is restored when the CDW amplitude, detected by Resonant Inelastic X-ray scattering, is suppressed. This observation points towards an intimate connection between the onset of CDW and the departure from $T$-linear resistivity in underdoped cuprates. Our results illustrate the potential of using strain control to manipulate the ground state of quantum materials.

**Main text**: Cuprate high temperature superconductors (HTS) belong to a class of materials where strong correlations play a fundamental role, and whose unconventional properties might require abandoning traditional concepts of solid-state physics for a proper description (*1*). The "strange metal" phase of HTS is one of the most striking manifestations of the strong correlations, and at optimal doping manifests as a linear temperature dependence of the resistivity that persists to the lowest *T* when superconductivity is suppressed (*1,2,3,4*). This behavior is fundamentally different from that observed in more conventional metals (*3*), where a *T*-linear dependence of the resistivity is found only at high temperatures where phonon scattering dominates the transport. The *T*-linear resistivity is also found in other strongly correlated systems, including pnictides (*5*) and magic angle bilayer graphene (*6*).

Recent developments attempt to model this behavior by considering that the scattering time approaches the fundamental Planckian limit defined by $\tau=\hbar/k_BT$, where $\hbar$ and $k_B$ are the reduced Planck and Boltzmann constants), irrespective of the nature of the scattering process (*4,7*). Ordinary metals, based on local interactions among quasiparticles, cannot thermalize on such a short time scale; to reach the Planckian limit one would require every particle to be entangled with every other particle in the system.

In cuprates, the *T*-linear resistivity is lost for doping above and below the optimal one. In the overdoped regime, the recovery of an almost $T^2$ dependence of the resistivity, typical of a Fermi liquid, is a consequence of the increased screening of the electron-electron interactions caused by a higher charge carrier density. In the underdoped regime the deviation from the *T*-linear behavior happens at temperatures close to *T**, known as the pseudogap temperature, where states are missing at the Fermi energy (*8*). In the pseudogap region, the HTS phase diagram also hosts a plethora of intertwined electronic local ordering phenomena breaking rotational/translational symmetry (*1,9,10,11,12,13*); Charge Density Wave (CDW) order (*11*) is the most prominent one. The association between the departure from the *T*-linear resistivity and the occurrence of the pseudogap phenomenon has long been speculated. However, there is no consensus on the physics at play and, more importantly, on the causality hierarchy among pseudogap, local orders, and strange metal phenomenology (*14,15*). This is because the strange metal exhibits its most salient features in transport and its connection to spectral signatures is elusive (*16*). More specifically, the challenge is to disentangle the various possible mechanisms leading to the breakdown of the *T*-linear resistivity, such as the occurrence of the pseudogap and the appearance of local orders, like CDW. One way to address this challenge is to tune the local properties of underdoped HTS. In particular, the CDW can be strongly modified under pressure (*17,18*), strong magnetic fields (*19*) and strain in crystals (*20*) and thin films (*21*).

To tune the ground state in thin films of the cuprate material $YBa_2Cu_3O_{7-\delta}$ (YBCO), we use the geometric modification of its unit cell under the strong strain induced by the substrate. We show that the *T*-linear resistivity dependence is completely recovered (down to the superconducting critical temperature $T_c$) along the *b*-axis when the CDW, detected by Resonant Inelastic X-ray scattering (RIXS), is suppressed along the *a*-axis.

### Thin film devices

The films we use span a wide range of hole-doping $p$, going from the strongly underdoped ($p \approx 0.10$) up to the optimally doped ($p \approx 0.17$) regime (*22,23*). The strain was modified both by changing the substrate and by varying the film thickness $t$ in a range from 50 nm down to 10 nm. We used (110) oriented MgO and 1° vicinal angle (001) SrTiO$_3$ (STO) substrates to grow untwinned YBCO films (*24*). When the YBCO thickness is reduced to a few unit cells ($t = 10$ nm), films grown on MgO are characterized by a significant elongation of the $b$-axis and contraction of the $c$-axis, with the total volume of the unit cell unchanged compared to relaxed systems (Fig. 1A). For films grown on vicinal cut STO, the YBCO unit cell is instead almost thickness independent (Fig. S1).

Figure 1B shows a typical device used to measure the resistivity as a function of temperature. The devices are patterned at an angle $\phi$ with respect to the YBCO [100] direction by using a carbon mask in combination with electron beam lithography and Ar$^+$ ion milling (*22,23,25*).

### Angular dependence of in-plane resistivity

Figure 2A shows the temperature dependence of the resistivity $\rho$, measured in two devices oriented along the YBCO $a$- and $b$-axis ($\phi = 0°$ and 90° respectively) realized in an underdoped ($p = 0.11$) 50 nm thick film grown on an MgO substrate. The resistivity anisotropy ratio at $T = 290$ K, defined by $\rho_a(290\ \text{K})/\rho_b(290\ \text{K})$ is around 1.2, a value fully compatible with the level of hole doping (*26*). The temperature $T_L$, estimated as the temperature below which the resistivity normalized to 290 K deviates by 1% from the linear fit, is approximately the same for both devices, and it is very close to the pseudogap temperature $T^*$ of single crystals at comparable doping measured using spectroscopic techniques (*8,27*). For this film thickness we therefore have that $T_L^a \approx T_L^b \approx T^*$.

Figure 2B shows analogous data for two devices patterned on a 10 nm thick film on MgO (grown with the same deposition conditions as the 50 nm thick film) with comparable doping $p$ as the one in Figure 2A. The superconducting critical temperature $T_c$ at this reduced thickness remains almost unchanged (*28*) compared to 50 nm thick films. However, the overall $\rho(T)$ behavior is very different. We observe three striking features, namely 1) the in-plane anisotropy of the resistivity $\rho_a(290\ \text{K})/\rho_b(290\ \text{K})$ is much larger compared to the 50 nm thick film, 2) the slopes $\gamma_{a,b}$ of the high temperature linear resistivity along the $a$- and $b$-axis are very different and 3) the resistivity along the $b$-axis $\rho_b(T)$ has a much broader temperature range of linearity, as is clearly highlighted by subtracting from $\rho_b(T)$ the linear fit $\rho_L(T) = \rho_0 + \gamma_b T$ of the high temperature resistive behavior (Fig. 2B, inset). Here $\rho_0$ and $\gamma_b$ are respectively the intercept at $T=0$ and the slope of the high temperature linear dependence. At $p = 0.14$ and $p = 0.147$ (Figs. S4C and 2C), for 10 nm thick films we find that the $T$-linear behavior is completely recovered down to the superconducting transition. This finding is crucial because it indicates that the "strange metal" behavior is restored in ultrathin underdoped YBCO. What are the conditions for this to happen?

The most prominent structural effect we encounter by reducing the thickness of the films is that the YBCO *b*-axis expands, whereas the *a*-axis is only slightly modified; at the same time the total volume of the cell remains unaltered, as a consequence of a *c*-axis contraction. One of the effects of the strain is therefore to increase the orthorhombicity of few-nm-thick films. For 10 nm thick underdoped films ($p \approx 0.12$) the values of the lattice parameters *a* and *b* are similar to those of YBCO with a much higher doping (*29*), close to the optimal doping. However this effect cannot explain the anomalously high anisotropy ratio $\rho_a(290\ \text{K})/\rho_b(290\ \text{K})$ that we observe (see Fig. 2B), a value that in an optimally doped YBCO crystals is mainly related to the presence of CuO chains (*26*). Indeed within a simple tight binding description, an increase of ~0.02 Å in the *b*-axis lattice parameter, as we observe upon reducing the film thickness from 50 nm to 10 nm (Fig. 1A), would reduce the corresponding hopping parameter between neighbor sites in the *b* direction by only ≈1% (*30*). Moreover, the resulting (weak) modification of the electronic structure would induce the opposite anisotropy in the *a*- and *b*-axis resistivities, namely $\rho_b > \rho_a$ (because of the larger *b*-axis value), compared to what we experimentally observe. This rules out the increased orthorhombicity as a possible, direct, explanation of the anisotropy we observe.

The very different slopes of $\rho_a(T)$ and $\rho_b(T)$ hints to the physics at play in the 10 nm thick films. Following the Boltzmann transport theory, the conductivity is given by

$$\sigma_{a,b}(T) = 2e^2 \sum_k \frac{v_{F,a,b}^2}{\Gamma(k)} \{-n'_F\}, \qquad (1)$$

where $v_{F,a,b}$ is the Fermi velocity, $\Gamma(k)$ the *k* dependent scattering rate and $n'_F$ the derivative of the Fermi distribution. Consequently, the ratio $\Gamma(k)/v_{F,a,b}^2$ determines the slope $\gamma_{a,b}$ of the temperature dependent film resistivity. From the experimental data we have $\gamma_a \gg \gamma_b$ (at $p = 0.11$ $\gamma_a$ = 4.9 μΩcm/K and $\gamma_b$ = 2.1 μΩcm/K) which indicates that $v_{F,b} \gg v_{F,a}$ and, therefore, that the Fermi surface is strongly anisotropic already at room temperature. This is illustrated in Figs. 2, D and E, which show respectively the typical isotropic Fermi surface for the cuprates (where the contribution of the chains is neglected) and an anisotropic distorted Fermi surface, compatible with our experimental resistivity anisotropy. From Eq. (1) one may argue that also anisotropic elastic scattering processes, e.g. due to small-angle scattering from impurities between $CuO_2$ planes (*31*), can account for the observed resistivity anisotropy. However, in this case $\Gamma(k)$ would be determined by the local density of states (*31*), i.e. $\Gamma$ would still be proportional to $1/v_F$. Therefore, only an anisotropy of the Fermi velocity, which breaks the $C_4$ symmetry of the crystal, can account for the observed resistivity anisotropy.

A Fermi surface of the type shown in Fig. 2E can be the consequence of a strong electronic nematicity in the system. Such a state has been extensively investigated from a theoretical point of view (*32,33,34,35,36*) and it has been experimentally found, by in plane resistivity anisotropy measurements, in tetragonal $La_{2-x}Sr_xCuO_4$ films (*37*). We speculate that in the 10 nm thin films the strain-induced distortion of the cell plays a fundamental role in stabilizing a nematic ground state already at room temperature.

The wider $T$-linear behavior of $\rho_b(T)$ in ultrathin films is not observed in YBCO on vicinal angle STO substrates. Here $T_L$ does not change: it is the same along the $a$-axis and the $b$-axis and going from 50 nm to 10 nm thick films (see Fig. S3). This in agreement with the fact that the slopes $\gamma_{a,b}$ of $\rho_a(T)$ and $\rho_b(T)$ are comparable in 10 and 50 nm thick films, so on STO substrates the Fermi surface is not significantly modified by strain at reduced thicknesses.

We have therefore arrived at the main result of our paper: a specific strain in underdoped 10 nm thick YBCO films on MgO substrate induces a nematic state that modifies the Fermi surface already at room temperature. But why would a distorted Fermi surface recover the $T$-linear resistivity behavior along the $b$-axis? The answer to this question comes from the study of the $\rho(T)$ dependence as a function of doping. Figure 3 shows $\rho_a(T)$ and $\rho_b(T)$ respectively as a function of the doping $p$ (Figs. 3, A and B) and of the angle $\phi$ (Figs. 3, C and D) for 10 nm thick YBCO films on MgO substrates (as a function of the thickness in Fig. S4).

The $(T)$ curves are rather conventional and in agreement with previous results (*22,26*). However, $\rho_b(T)$ (Fig. 3B) looks very different. As anticipated earlier, for $p \approx 0.14$ we have completely recovered the strange metal behavior: the $T$-linear dependence extends through the entire temperature range until superconductivity sets in. However the situation changes at lower doping: for $p \approx 0.10$, the extracted $T_L$ along $b$ is close to the value along the $a$-axis (see Figs. 2, C and D), and $\rho_b(T)$ shows a pronounced upturn at low temperature before the superconducting transition. The upturn of the $\rho_b(T)$, in our 10 nm thick YBCO films, is observed at higher doping compared to the $\rho_a(T)$ (Figs. 3, A and B); indeed, it appears already at $p = 0.13$. This upturn in the resistivity has been attributed to the loss of high mobility electron pockets because of the CDW order ending at $p \approx 0.08$ (*38*) and/or to the proximity to the antiferromagnetic Mott insulator through spin density waves (SDW) (*39,40,41*). The doping dependence of $(T)$, therefore, points towards a strong involvement of the CDW order in the phenomenology we observe.

### Resonant inelastic x-ray measurements of CDW

To characterize the CDW order in the YBCO films we have used resonant inelastic x-ray scattering (RIXS) at the Cu $L_3$ edge (~930 eV) (*23*). We investigated films with a thickness of 10 and 50 nm grown on both MgO and STO at the doping $p \approx 0.125$ where the intensity of the CDW is strongest (*11,42*). To isolate the contribution of the CDW we measured RIXS spectra at $T = 70$ K, i.e. close to $T_c$, where the CDW signal is maximized, and at $T = 200$ K, a temperature where the CDW contribution is negligible. The CDW peak has been explored along $a$ and $b$ with two orthogonal cuts along the $H$ and the $K$ directions of reciprocal space, centered around the wave vector of the CDW $q_c^{CDW} = (H_{CDW}, K_{CDW})$. Figures 4, A and B, show, for a 50 nm thick film on MgO, the quasi-elastic component of the RIXS spectra along the $a$-axis and $b$-axis, as a function of $H$, at $T = 70$ K and $T = 200$ K. Along both directions, at $T=200$ K only a broad peak is present

(see red regions in Figs. 4, A and B); at $T = 70$ K, the signal is given by the sum of a broad peak (similar to that measured at high temperature) and of a narrow peak. This narrow, temperature-dependent, peak is a signature of CDW; the broad, almost temperature-independent, peak is instead a signature of short-ranged charge density fluctuations (CDF) (*43*), i.e. charge modulations, precursors of CDW, which are present in the phase diagram at any temperature and in a very broad doping range, including the overdoped region. Figures 4, C and D, report the same measurements, as in Figs. 4, A and B, but for a 10 nm thick film on MgO. Here, along the *b*-axis (see Fig. 4D) both the CDW and the CDF peaks are very similar to those measured in the thick film. Along the *a*-axis (see Fig. 4C), the situation is instead dramatically different. The broad-in-$q_{//}$ CDF peak is unchanged with respect to the 50 nm thick sample; in contrast, the narrow CDW peak, emerging at lower temperature, is almost negligible. This occurrence has been verified on the same sample, measuring along the *K* direction (see Fig. S5), and on other ultrathin films on MgO with different *p* doping (see Fig. S6). It should be noted that our films are not perfectly untwinned (the untwinning degree is ≈85% for films grown on MgO, ≈90% for films grown on STO). The small temperature dependent CDW signal measured along the *a*-axis in the 10 nm films on MgO can be attributed to twinned domains. Indeed, once the twinning is taken into account, the actual CDW signal becomes effectively negligible in the *a*-axis direction. We conclude that in 10 nm thick films on MgO the CDW is unidirectional and directed along the *b*-axis, whereas the CDW in YBCO grown on STO is thickness-independent (Fig. S7). Strain induced modifications of CDW have been also recently observed in YBCO single crystals, where the in-plane uniaxial compression along either *a* or *b* gives rise to an enhancement of the CDW in the orthogonal direction (*44*).

**Discussion**

Our results are in line with the theoretical predictions of unidirectional CDW in the presence of nematicity (*34,35*), as a consequence of the modified Fermi surface. Our experiment additionally shows a clear correlation between the *T*-linear dependence of $\rho(T)$ and the charge order: the recovery of the *T*-linear resistivity along the *b*-axis is associated with the suppression of the CDW along the YBCO *a*-axis. In this scenario, only CDF survive in the system at any temperature, which might be relevant both for Planckian metal (*45*) and marginal Fermi liquid theories (*46*) of the strange metal.

The strong correlation between the CDW and the breakdown of the *T*-linear behavior is further supported by the co-incidence of the onset temperatures for the two phenomena. Figure S8 shows the temperature dependence of the CDW order for our 50 nm and 10 nm thick films on MgO and STO substrates. Within the experimental error, $T_{\mathrm{CDW}}$ (for $p = 0.125$) is the same as the temperature at which the resistivity departures from the linearity $T_L^a$ and the same as the pseudogap temperature *T*\* (taken from literature (*8, 27*)). This is illustrated in Figure 5 which shows a revised phase diagram for the 10 nm thick films. Our RIXS measurements therefore require the revision of the common belief that the CDW is a low temperature phenomenon happening at temperatures significantly below $T_L$ and *T*\* as indicated in early works (*11,42*). The coincidence between *T*\*,

$T_L$ and $T_{CDW}$ is a common feature of doping above the 1/8 where the CDW order is strong, as also supported by recent RIXS data (*43*), which are included for completeness in Figure 5.

The quality of our samples allows us to exclude that possible structural changes of the CuO chains along the *b*-axis in 10 nm films might have a role in the phenomenology we have observed. For overdoped films, where the conductivity of the chains causes an upward deviation from linearity in $\rho(T)$, we observe the same $\rho_b(T)$ dependence for 50 nm and 10 nm films and very similar values of $T_L$ (Fig. S2), which excludes CuO chain modification effects in our very thin films.

However, there is still an issue to be addressed: why – in our films – the disappearance of the CDW along the YBCO *a*-axis causes a recovery of the *T*-linear dependence of the resistivity along the *b*-axis (see Figs. 2B and 4C)? It is not uncommon in other compounds that the occurrence of a unidirectional CDW phenomenon affects the transport properties in a direction orthogonal to the CDW *q*-vector (*47*). In our experiment, the reason can be linked to the nematic Fermi surface for very thin films. The generic Fermi surface (see Fig. 2D) supports two scattering processes along the *b*-axis associated with the quasi-nesting properties of the CDW *q*-vector: one within the Brillouin zone and the other connecting two adjacent Brillouin zones at $(\pi,0)$ (see Fig. S9A). For a conventional Fermi surface the scattering rates of these two processes are mostly comparable (*23*). However, for a distorted Fermi surface of Fig. 2E the region around the $(\pi, 0)$ point has a much higher density of states compared to the unperturbed one (see Fig. S9). As a consequence, CDW scattering rate involving two adjacent Brillouin zones becomes more than four times larger than the scattering rate within the same Brillouin zone (*23*). Following the Boltzmann equation formalism, this strongly affects the resistivity along *a*-axis, while leaving the resistivity along the *b*-axis unaltered.

Finally, it is important to comment on the role of the pseudogap in the departure of the *T*-linear resistivity in underdoped cuprates. From the slopes of $\rho(T)$ along the *a*- and *b*-axis $\gamma_{a,b}$ we infer that the Fermi surface is anisotropic at any doping (the ratio $\gamma_a/\gamma_b$ is almost doping independent). For $p \leq 0.1$, we observe that $T_L$, which is at that doping level much higher than $T_{CDW}$ and comparable to $T^*$ (taken from literature (*8,27*)) is isotropic ($T_L^a \approx T_L^b$). This hints at a pseudogap that is isotropic on a distorted Fermi surface. Such an isotropic pseudogap cannot explain the otherwise anisotropic $T_L$ we observed at higher doping ($p > 0.1$). This allows us to conclude that the pseudogap is not the main effect responsible for the departure of the linearity when the CDW order is strong ($p > 0.11 - 0.12$), which is the range of doping in our experiment. However, one cannot exclude a major influence of the pseudogap on the transport at lower doping when $T^*$ becomes much higher (close to room temperature and beyond) than $T_{CDW}$.

**Acknowledgments:** The authors acknowledge enlightening discussions with M. Grilli, C. Di Castro, M. Moretti. They also thank D. Betto, for initial tests at ID32 on ultrathin YBCO films, and L. Martinelli for helping with complementary RIXS measurements. This work was performed in part at Myfab Chalmers. The RIXS experimental data were collected at the beam line ID32 of the European Synchrotron (ESRF) in Grenoble (France) using the ERIXS spectrometer designed jointly by the ESRF and Politecnico di Milano.

**Funding:** This work has been supported by the European Union's projects NANOCOHYBRI (Cost Action CA16218) and OXiNEMS (Horizon 2020, Grant Agreement No. 828784), by the Knut and Alice Wallenberg foundation (project No. 2014.0102) and by the Swedish Research Council (VR), under the projects 2015-04368 (U.G.), 2018-04658 (F.L.), 2020-04945 (R.A.) and 2020-05184 (T.B.). G.S. acknowledges support from the the Deutsche Forschungsgemeinschaft under SE 806/19-1. G.G. and S.C. are partly supported by Italian Ministry of University and Research (PRIN Project No. 2017Z8TS5B). S.C. also acknowledges financial support from the University of Rome Sapienza, through the projects Ateneo 2018 (Grant No. RM11816431DBA5AF), Ateneo 2019 (Grant No. RM11916B56802AFE), Ateneo 2020 (Grant No. RM120172A8CC7CC7).

**Author contributions:** F.L., E.W., R.A., and T.B. conceived and designed the experiments. E.W., R.A., and E.T. grew the samples. E.W. performed the transport measurements. R.A., E.W., F.L., T.B., M.R., R.F., G.G., and N.B.B. performed the RIXS measurements. R.A., G.G., and L.B. analysed and interpreted the RIXS experimental data. F.L., E.W., R.A., and T.B. interpreted the transport measurements with the theoretical insights of U.G.. G.S. and S.C. performed the modelling of the experimental data and the theoretical calculations. F.L., R.A., E.W., U.G. and T.B. wrote the manuscript with contributions from all authors.

**Competing interests:** The authors declare no competing financial interests. G.G. is member of the Science Advisory Council of the ESRF.

**Data and materials availability:** All experimental data shown in the main text and in the supplementary materials are accessible at the Zenodo repository (*48*).

**Supplementary Materials:**
Materials and Methods
Supplementary Text
Fig S1 – S9
References (*49 – 64*)

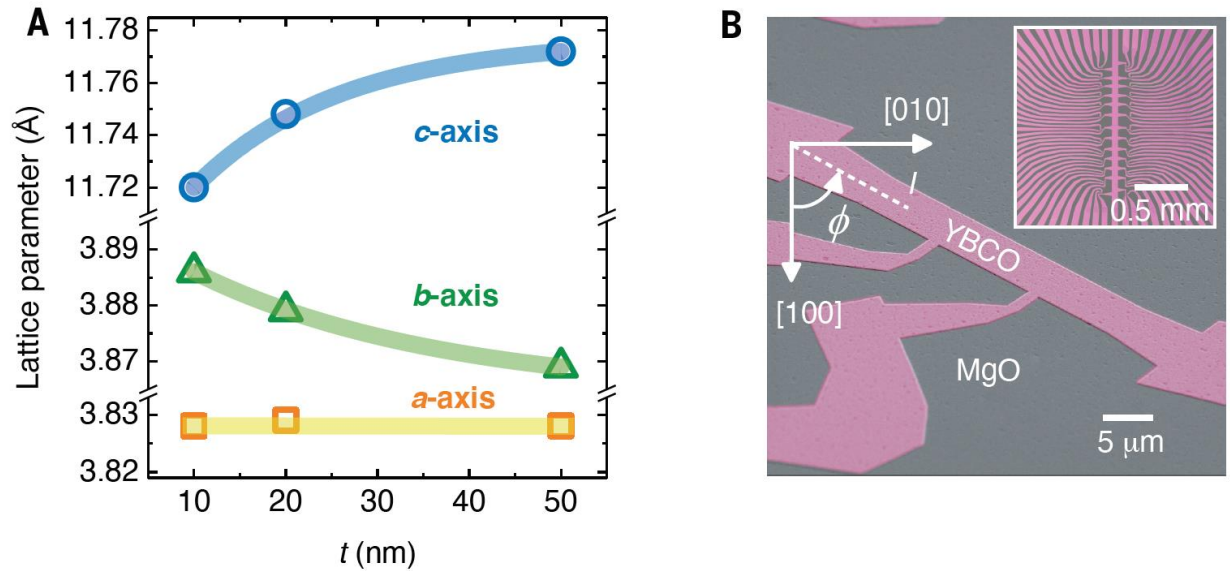

**Figures:**
**Fig. 1**: **Thickness dependence of the YBCO lattice parameters for films grown on MgO substrates.** (**A**) Lattice parameters (*a*- *b*- and *c*-axis indicated by squares, triangles and circles respectively) of YBCO, measured by x-ray diffraction at 300 K, as function of the thickness of films with a hole-doping $p \approx 0.12$ grown on MgO. Thick lines are guides to the eye. (**B**) False colour scanning electron microscope image of a typical device used to measure the resistivity $\rho$, with the current $I$ flowing at an angle $\phi$ with respect to the YBCO [100] direction (*a*-axis). Inset: Overview of the patterned samples.

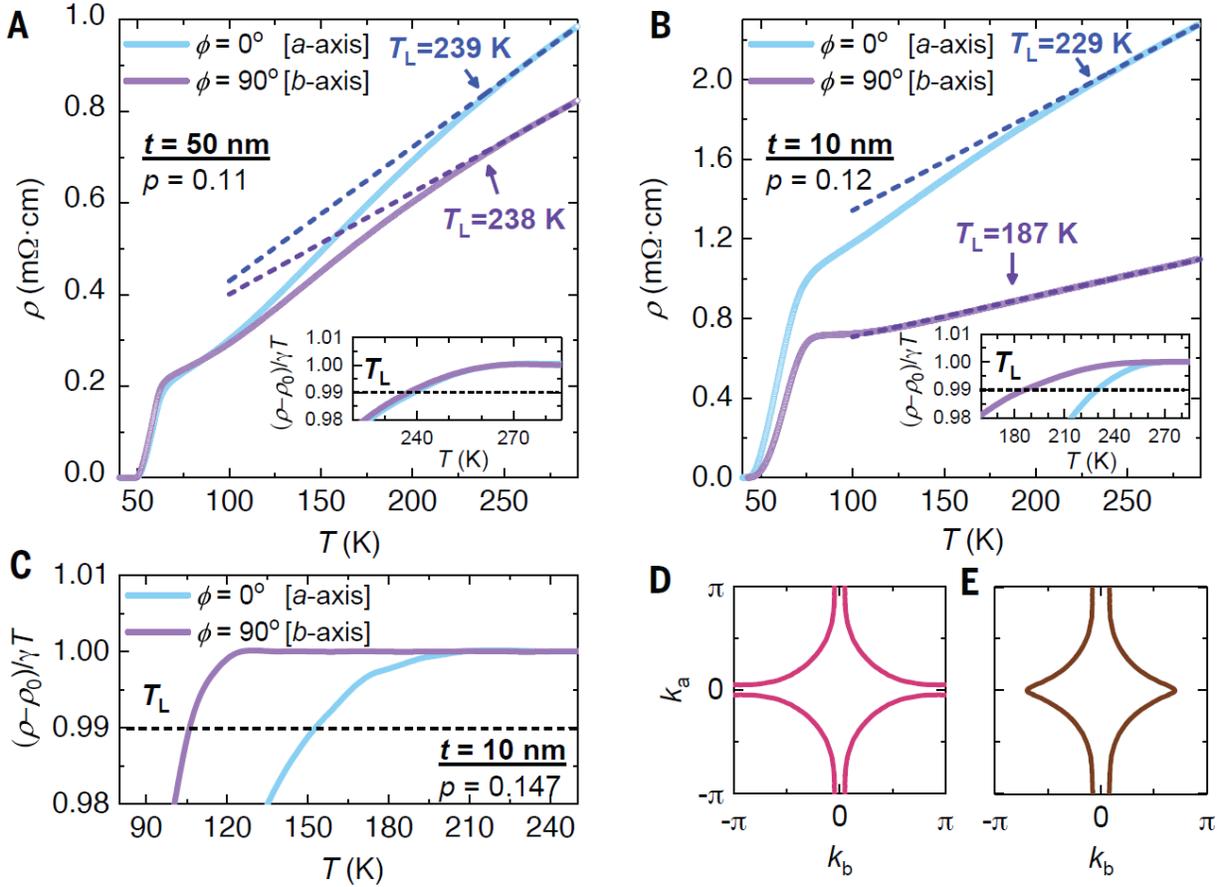

**Fig. 2: Angular dependence of the YBCO in-plane resistivity as a function of the thickness of films on MgO under strain.** (**A**) $\rho(T)$ of two devices, oriented along the *a*- and the *b*-axis directions, patterned on a 50 nm thick underdoped (*p*=0.11) film. At $T = 290$ K, $\rho_a/\rho_b = 1.2$. The dashed lines are the linear fits of the curves for $T > 260$ K. The inset shows the determination of $T_L$ which is the temperature where the resistivity normalized to 290 K deviates by 1% from the linear fit ($\rho_0$ and $\gamma$ are the coefficients of the fit). (**B**) $\rho(T)$ of two devices, oriented along the *a*- and the *b*-axis directions, patterned on a 10 nm thick underdoped (*p*=0.12) film. At $T = 290$ K, $\rho_a/\rho_b = 2.1$. (**C**) Determination of $T_L$ in two slightly underdoped devices, oriented along the *a*- and the *b*-axis directions, patterned on a 10 nm thick slightly underdoped (*p*=0.147) film. $\rho(T)$ along the *b*-axis is linear down to the onset of superconducting fluctuations around $T = 100$ K. (**D**) Simplified model of the typical Fermi surface of cuprates (here we disregard the influence of the CuO chains). For this kind of Fermi surface, the resistivity should be isotropic in the copper oxide planes. (**E**) Hypothetical anisotropic model Fermi surface that is compatible with the anisotropic transport of the 10 nm thick devices.

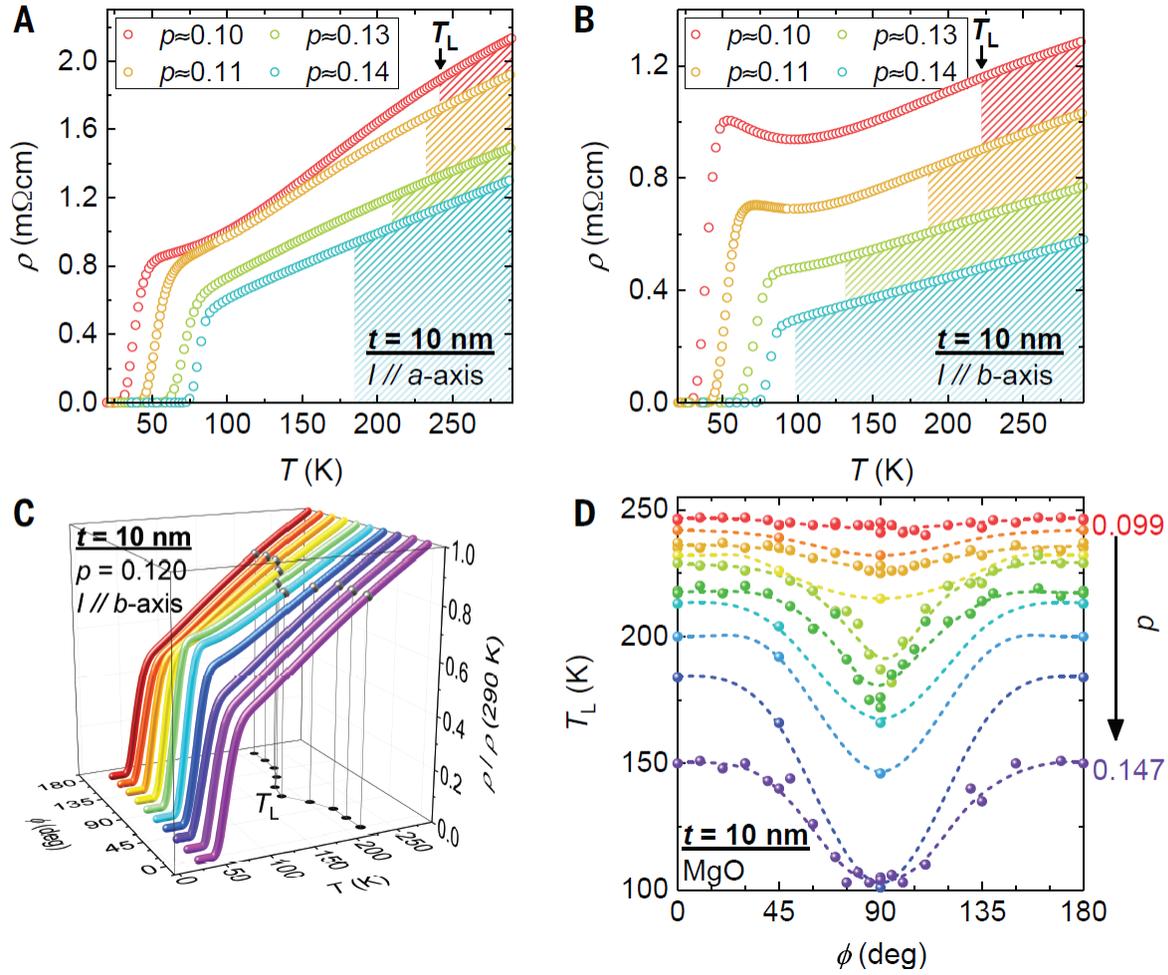

**Fig. 3: Angular dependence of the in-plane resistivity of the 10 nm YBCO on MgO as a function of the oxygen doping.** (A) $\rho(T)$ measured along the *a*-axis direction of 10 nm thick films, with different doping levels, grown on MgO substrates. $T_L$ is extracted as described in Fig. 1. The dashed regions represent the temperature range above $T_L$ where at each $p$ the $T$-linear resistivity regime occurs. (B) Same as panel (A), but $\rho(T)$ is measured along the *b*-axis. (C) $\rho(T)$, normalized to its value at $T$=290 K, for a 10 nm thick film ($p$=0.120) on MgO is shown as a function of $\phi$, i.e. with respect to the *a*-axis direction. For each angle, the temperature $T_L$ has been extracted (black spheres and dots). $T_L$ is approximately constant as a function of $\phi$, except around $\phi$=90°, where it exhibits a suppression. (D) $T_L$ values as a function of $\phi$ for 10 nm thick films with different doping levels ($p \approx$ 0.099, 0.103, 0.108, 0.117, 0.118, 0.120, 0.123, 0.134, 0.140, 0.147).

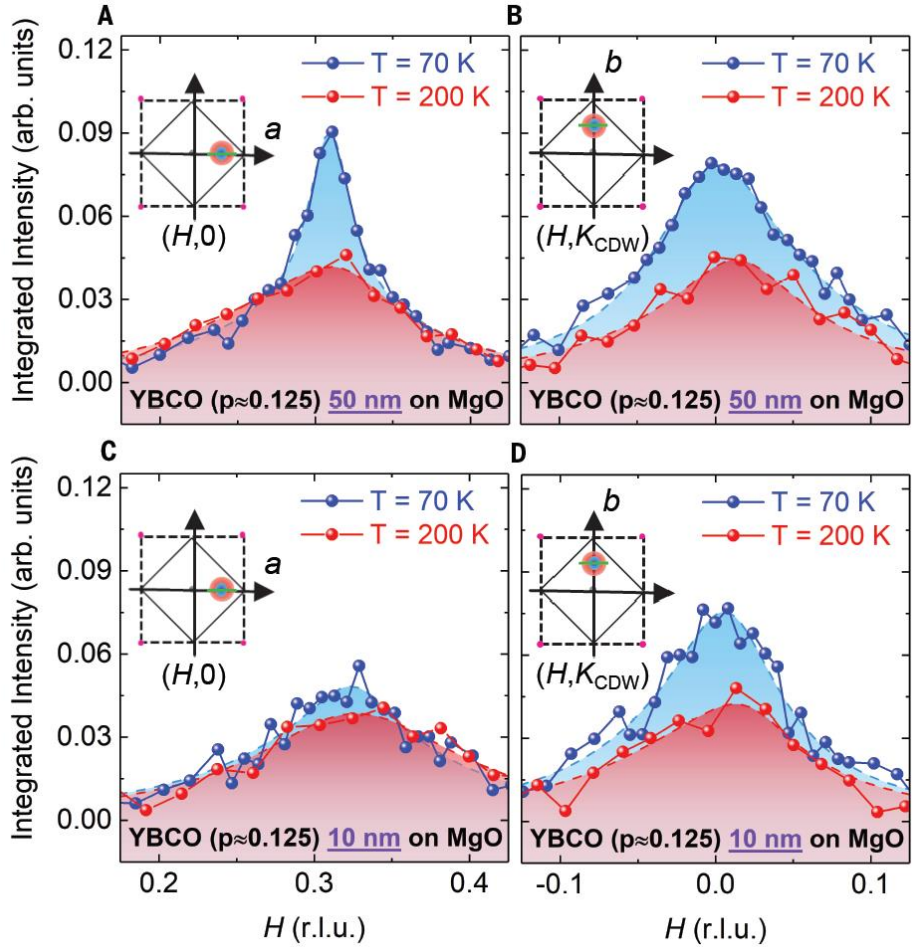

**Fig. 4**: **Thickness dependence of the CDW in YBCO thin films.** **(A)-(B)** Quasi-elastic scans measured at $T = 70$ K and $T = 200$ K on the 50 nm thin film on MgO along the $(H,0)$ direction, i.e. $a$-axis (A), and the $(H,K_{CDW})$ direction, i.e. $b$-axis (B). **(C)-(D)** Same as panels (A)-(B), but on a 10 nm thick sample. The measurements have been performed along the $(H,0)$ direction, i.e. $a$-axis (C), and the $(H,K_{CDW})$ direction, i.e. $b$-axis (D). In the 10 nm thick sample, the CDW intensity along the $a$-axis is almost negligible. If we take into account the percentage of twin domains ($\approx 15\%$) present in our films, we conclude that in few-unit-cell thick films on MgO the CDW is unidirectional along the $b$-axis. The green line in the inset of each figure shows the direction of the scan relative to the CDW peak in reciprocal space.

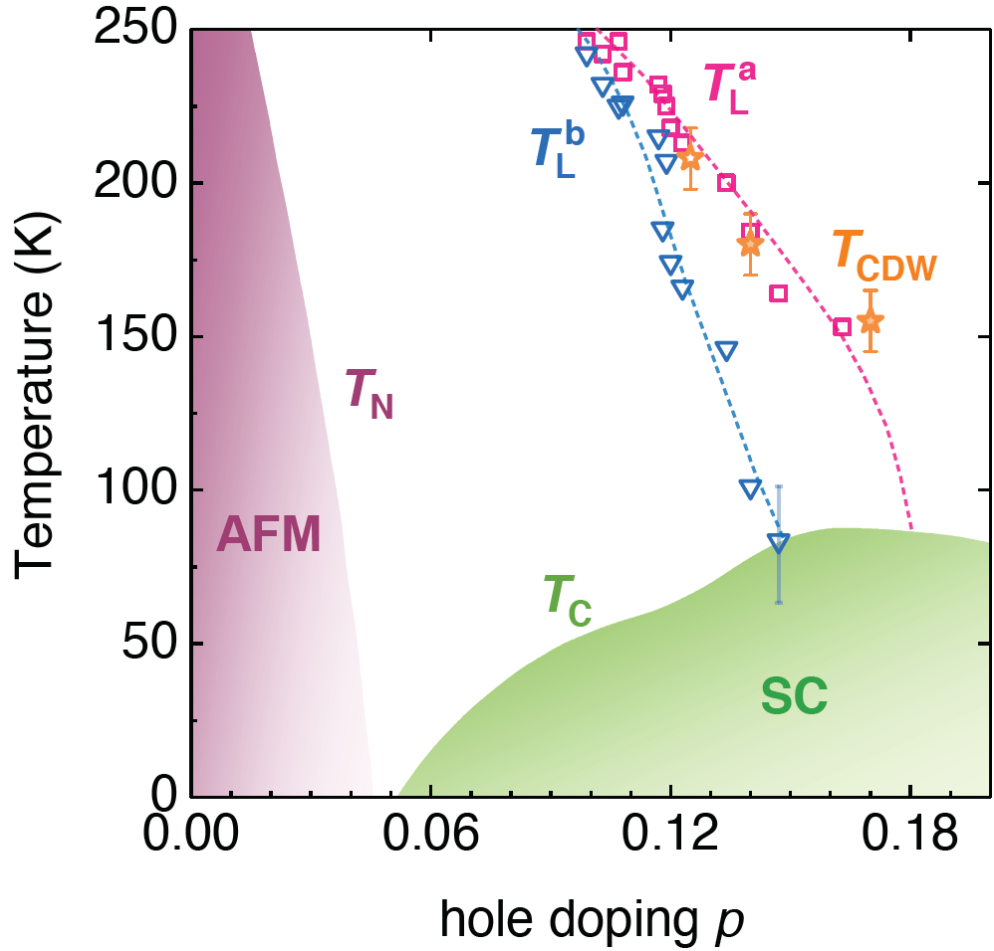

**Fig. 5: Intertwined phases in the phase diagram of YBCO ultrathin films.** The antiferromagnetic (AFM) and superconducting (SC) states are present below $T_N$ and $T_c$, respectively (the $T_c$ dome is derived from Ref. 22). $T_L$ serves as a crossover line between the quantum-critical region of the phase diagram exhibiting strange metal properties (on the right side of $T_L$) and the non-critical region, where the linearity of the resistivity versus temperature is lost (on the left side of $T_L$). Along the $a$-axis $T_L=T_L^a$ (pink squares) is close to the pseudogap temperature $T^*$ as determined by other techniques, whereas along the $b$-axis (blue triangles), $T_L$ is suppressed to lower temperatures. The error bar on $T_L^b$ highlights the uncertainty of determining the correct value owing to proximity to $T_c$. The lines are guides to the eye. $T_{CDW}$ (stars) is the onset temperature of the CDW, as determined by resonant inelastic X-ray scattering. The empty star is related to the doping level investigated by RIXS in this manuscript (for 10 nm thick films, revealed only along the b-axis), whereas the filled stars are from (43).

# Supplementary Materials for

# Restored strange metal phase through suppression of charge density waves in underdoped YBa$_2$Cu$_3$O$_{7-\delta}$

Eric Wahlberg, Riccardo Arpaia[*], Götz Seibold, Matteo Rossi, Roberto Fumagalli, Edoardo Trabaldo, Nicholas B. Brookes, Lucio Braicovich, Sergio Caprara, Ulf Gran, Giacomo Ghiringhelli, Thilo Bauch, Floriana Lombardi[*]

[*] Corresponding author. E-mail: floriana.lombardi@chalmers.se (F.L.); riccardo.arpaia@chalmers.se (R.A.)

**This PDF file includes:**

    Materials and Methods
    Supplementary Text
    Figs. S1 to S9
    References (*49 – 64*)

**Materials and Methods**

Samples

The YBa$_2$Cu$_3$O$_{7-\delta}$ (YBCO) films have been deposited by pulsed laser ablation on 5 × 5 mm$^2$ (110) oriented MgO and 1° vicinal angle (001) SrTiO$_3$ substrates. To favor the growth of untwinned films, an anisotropic strain is needed, preventing the random exchange of the *a* and *b* axis parameters. On STO this is achieved by the formation of terraces on the surface, therefore the substrate vicinality is the critical control parameter (*49*); on MgO similar role is played by facets, which are induced by a 5-hours-long in-situ annealing at 790 °C, done before the deposition (*24*). Details on the sample growth have been previously reported (*22*). In particular, after the deposition, the films have been slowly cooled down at an oxygen pressure varying in a range between 30 μbar and 600 μbar, to span most of the underdoped side of the YBCO phase diagram.

Benchmark of the quality of the films is the phase diagram we can build from transport measurements (*22*), which is in qualitative and quantitative agreement with that of YBCO single crystals (*50*). In particular, the doping *p* for each film cannot be directly determined, since a precise estimate of the oxygen content *n* and of the degree of oxygen ordering in the chains can be determined, although with several challenges, only in single crystals. Therefore, using a method successfully used for crystals (*51*), we have indirectly determined *p* by the knowledge of $T_c$, in combination with the *c*-axis length obtained via XRD: for films with thickness larger than 30 nm, where the *c*-axis length is strain-, therefore thickness-, independent, the result is that the doping *p*, for every value of critical temperature $T_c$, is in good agreement with the doping *p* measured in single crystals (in this paper, we refer in particular to Refs. *42, 51*).

Details on the growth and characterization of our few-unit-cell thick films, the 10 nm thick ones, to which we refer as "ultrathin" in the text, have been previously reported (*28*). These films are characterized by high crystallinity and by properties representative of the YBCO bulk. In particular, for a fixed doping level, the critical temperature $T_c$ is – within few Kelvin – identical to that of thicker, 50 nm films. The small discrepancy, which becomes more significant when decreasing even more the thickness, is due to the increased broadening of the superconducting transition, explained in terms of the Kosterlitz-Thouless transition in systems which tend to become 2D (*28, 52*).

Determination of the thin film lattice parameters

The lattice parameters of the films have been determined by XRD analysis using a Panalytical X'pert PRO Materials Research 4-axis diffractometer with 4-bounce Ge220 monochromator and PIXcel detector. The *c*-axis parameter has been determined by symmetric 2θ-ω scans, from the angular position of the (00n) YBCO peaks. To find out the length of the *a* and *b* axis parameters, as well as the twinning state of the films, we have measured the asymmetrical (038)-(308) reflections of YBCO via 2θ-ω maps (see Fig. S1). For films grown on MgO the untwinning degree is in the range 82-85%, while it exceeds 90% for films grown on vicinal angle STO substrates.

Transport measurements

Transport measurements have been carried out using a Quantum Design PPMS, with a 4-point measurement scheme. The devices, through which the current flows, are wires, whose width is in the range between 5 μm and 15 μm (see Fig. 1B). The voltage probes, orthogonal to the wires, are 10 μm distant. The nanopatterning procedure we have used is based on an amorphus carbon mask, in combination with electron beam lithography and a gentle ion milling. We have routinely used this procedure in recent years to achieve structures of several cuprate compounds, including YBCO, that present properties representative of the bulk superconducting material and unaffected by the nanopatterning procedure, down to 50 nm (*25,53,54,55*).

The resistivity of the wires has been measured in the temperature range between 10 K and 290 K. For each measured curve, we have determined the critical temperature $T_c$ at 90% of the normal resistivity just above the transition. $T_L$ is instead defined as the temperature where the resistivity, normalized to 290 K, deviates by 1% from a linear fit at high temperatures.

RIXS measurements

The RIXS measurements have been taken at the beamline ID32 of the European Synchrotron Radiation Facility (ESRF) in Grenoble using the high-resolution ERIXS spectrometer (*56*). The YBCO films, which do not require any particular surface preparation prior to the measurements, were mounted on a 6-axis in-vacuum Huber diffractometer. The energy of the incident X-ray beam has been tuned to the maximum of the Cu $L_3$ absorption peak, around 931 eV. The instrumental energy resolution was set at 55 meV, determined as the full width at half maximum of the non-resonant elastic scattering from silver paint. Details of the experimental geometry has been already shown in previous papers (*57,58*): X-rays are incident on the sample surface, which is normal to the YBCO *c*-axis, and are scattered by an angle $2\theta$. The momentum transfers are given in units of the reciprocal lattice vectors $a^* = 2\pi/a$, $b^* = 2\pi/b$, $c^* = 2\pi/c$. We have worked at $2\theta = 149.5°$ in order to get $|q| = 0.91$ Å$^{-1}$, which allowed us to cover the whole first Brillouin zone along the [100] and [010] directions (0.5 r.l.u. ≈ 0.81 Å$^{-1}$). By rotating the samples around the $\theta$ and $\chi$ axes, without moving $2\theta$ (in the hypothesis – supported by previous experiment on cuprates - of a small interlayer coupling, therefore of a weak dependence of the charge order on $L$), we have changed the in-plane wave vector component $q_{//}$, measuring respectively the *H* scans and the *K* scans. At each $q_{//}$ value, the RIXS spectrum has been measured in 2.5 minutes (sum of individual spectra of 15 seconds) for the 50 nm thick films and 7 minutes (sum of individual spectra of 60 seconds) for the 10 nm thick films. The zero-energy-loss line has been determined by measuring a RIXS spectrum, dominated by the elastic contribution, on a 50 nm Au frame sputtered on the edges of the films. The quasi-elastic intensity has been determined by the integral in a range ±100 meV around the zero-energy-loss line. The RIXS spectra presented in Fig. 3 have been normalized to the integral of the inter-orbital *dd* excitations, in the range [-3 eV, -1 eV]. We have thoroughly explored four YBCO films: two on MgO (with thickness $t = 10$, 50 nm) and two on STO (with thickness $t = 10$, 50 nm). All the films are characterized by a doping level $p \approx 0.125$ and $T_c$ varying in the range between 65 and 70 K. For each film, we have explored the CDW peak along both the [100] and the [010] directions: in the former case, the *K* scan (as in Figs. S5A, S5B, S7C and S7G) is performed at a value of *H* coinciding with the position of the maximum of the narrow CDW peak, $H_{CDW}$; in the latter case, the *H* scan (as in Figs. 3B, 3D, S7B and S7F) is performed at a value of *K* coinciding with the position of the maximum of the narrow CDW peak, $K_{CDW}$. For each investigated film, the intensity and the correlation length of the CDW differ between the two in-plane crystal axis directions (see Fig. 3), as expected in untwinned YBCO (*42, 59*). The almost *T*-independent CDF contribution, we have used to single out the CDW, has shown up to be, after the first discovery in YBCO (*43*), a rather general characteristic of HTS cuprates (*60,61,62*).

We have measured some of the films, grown both on MgO and on STO, as a function of the temperature, determining an onset temperature for CDW $T_{CDW} \approx 208$ K ± 10 K (see Fig. S8). Therefore, in order to isolate the contribution of the CDW, we have measured for all the investigated samples RIXS spectra at $T = 70$ K $\approx T_c$, where CDW are strongest, and at $T = 200$ K, where the CDW contribution is negligible. The high temperature measurements provide therefore the bare, almost *T*-independent, CDF contribution. The CDW peak is finally determined subtracting the CDF from the low temperature measurements.

**Supplementary Text**

Distorted Fermi surface from the slopes of the high-*T* linear resistivity along the *a*- and *b*-axis

The reason for a strongly distorted Fermi surface is related to the very different behavior of the linear $\rho(T)$ along the *a*- and *b*-axis in the thin films as shown in Figure 2E of the main paper. The reduction of the film thickness from 50 nm to 10 nm predominantly induces an overall increase of the *a*-axis resistivity whereas the *b*-axis resistivity is much less affected, but more importantly the slopes of the linear parts become very different. This can be easily understood considering an anisotropy in the *a*- and *b*-axis Fermi velocities $v_F^{a,b}$ based on Boltzmann transport theory,

$$\sigma_{a,b}(T) = 2e^2 \sum_k \frac{v_{F,a,b}^2}{\Gamma(k)}\{-n'_F\} , \quad (S1)$$

where $v_{F,a,b}$ is the Fermi velocity along the *a* and *b* direction, respectively, $\Gamma(k)$ the *k* dependent scattering rate and $n'_F$ the derivative of the Fermi distribution. The conductivity is essentially determined by the sum of $\left(v_F^{a,b}\right)^2/\Gamma$ over the Fermi surface. It should be noted that an anisotropy of the elastic part of $\Gamma$, e.g., via small angle scattering from impurities between CuO$_2$ planes (*31*), is determined by the local DOS, i.e., $\Gamma \sim 1/v_F$. This means that finally only an anisotropy of the Fermi velocity (and hence of the Fermi surface), which breaks $C_4$ symmetry, can account for the observed resistivity anisotropy, and in particular for the very different slopes of the (*T*) and $\rho_b$(T), in our 10 nm thick films. In an effective model for the thin film Fermi surface, the *a*-axis Fermi velocity should therefore be reduced. Since the Fermi velocity $v_F$ is always perpendicular to the Fermi surface this amounts to deform the Fermi surface segments which are perpendicular to the $k_a$. In terms of a tight-binding model, and considering different hopping parameters $t_a$ and $t_b$ along the in-plane *a* and *b* direction, the dispersion relation within a common parametrization for cuprates is

$$E(k) = -2\, t_b \cos k_b - 2\, t_a \cos k_a + 4\, t' \cos k_a \cos k_b,$$

so the condition discussed above can be achieved by hopping parameters which obey $|t_a| < |t_b|$.

The origin of this anisotropy, which causes a deformed Fermi surface, is possibly related to a nematic state, as already demonstrated in other works (*34, 35, 37*).

Finally, we highlight here that – in contrast to a nematic ground state - the level of strain and the resulting increase in orthorhombicity of the 10 nm thick films on MgO cannot on its own justify the strong modification of the Fermi surface as presented in Figure 2E. First, within a simple tight binding description an increase of ~0.02 Å in the *b*-axis lattice parameter upon reducing the film thickness from 50 nm to 10 nm, would reduce the corresponding hopping parameter $t_b$ by about 1% only [see e.g. Ref. (*30*) for a parametrization of Slater-Koster integrals as a function of distance]. Second, the resulting (weak) modification of the electronic structure would induce the opposite anisotropy in the *a*- and *b*-axis resistivities compared to what experimentally observed, which instead can be explained by an increase of $t_b$ (or decrease of $t_a$).

Implications of unidirectional CDW order on the in-plane transport properties

In the following we use the Boltzmann transport equations to reproduce the observed in plane transport properties as a function of temperature (*63*)

$$\frac{1}{\rho_a} = \sigma_a = \frac{e^2}{4\pi^3 \hbar} \frac{2\pi}{d} \int_0^{2\pi} d\varphi \frac{k_F(\varphi)v_F(\varphi)\cos^2(\varphi-\gamma)}{\Gamma(\varphi)\cos(\gamma)} \quad (S2)$$

$$\frac{1}{\rho_b} = \sigma_b = \frac{e^2}{4\pi^3\hbar}\frac{2\pi}{d}\int_0^{2\pi} d\varphi \frac{k_F(\varphi)v_F(\varphi)\sin^2(\varphi-\gamma)}{\Gamma(\varphi)\cos(\gamma)} \quad (S3)$$

where $\varphi$ is the angle between the in-plane Fermi wave vector $k_F(\varphi)$ and $(k_b, 0)$, $v_F(\varphi)$ denotes the Fermi velocity, $\Gamma(\varphi)$ the scattering function and $\gamma(\varphi)$ is the angle between $k_F(\varphi)$ and $v_F(\varphi)$. Eqs (S2) and (S3) have been derived for a planar system with distance $d$ between the planes.

Our analysis is based on the following assumptions (besides the overall in plane transport anisotropy $\rho_a \neq \rho_b$):

1. At high temperatures the resistivities follow a $\rho \sim T$ behavior which implies an inelastic contribution to the scattering rate $\Gamma_{inel} \sim T$.

2. Below a temperature $T = T_L$ the resistivity deviates from the high temperature linear behavior in such a way that $\rho(T)$ falls *below* its linear extrapolation. This implies an effective *negative* contribution to $\Gamma_{inel}$ below $T_L$ possibly due to the onset of a paraconductive contribution mediated by CDW fluctuations since for our samples $T_L$ coincides with the onset of CDW scattering as observed in RIXS.

3. For the 10 nm films, the CDW is suppressed along the a-direction. The CDW scattering along $b$ induces a departure of $\rho(T)$ from linear behavior below $T_L$ along the a-direction.

We consider the following angle dependent scattering function

$$\Gamma(\varphi) = \Gamma_0 \, \eta(\varphi)$$

with $\Gamma_0$ the isotropic elastic scattering rate and

$$\eta(\varphi) = 1 + \eta_{lin}(T) + \eta_{cdw}(T, \varphi).$$

Here

$$\eta_{lin}(T) = \zeta \, k_B T \, atan\left(\frac{k_B T}{\hbar\omega_{FL}}\right)$$

describes the isotropic scattering linear in temperature above a Fermi liquid crossover energy scale $\omega_{FL}$ which in the following is set to $\omega_{FL} = 15$ meV (*46*). For the temperature dependence of the CDW scattering we take a quadratic behavior below $T_L$

$$\eta_{cdw}(T, \varphi) = -\beta(T - T_L)^2 \, \Theta(T_L - T)f(\varphi) \quad (S4)$$

with positive valued $\beta$, representing a negative contribution to the overall scattering rate. The observed CDW vector ($Q_b$) is compatible with scattering along the antinodal Fermi surface segments of adjacent Brillouin zones and between nodal segments in the same Brillouin zone. As shown below, the dominating scattering channel is along the antinodal Fermi surface segments. Hence, the angular CDW scattering function $f(\varphi)$, considering a unidirectional CDW order along the *b*-axis, is peaked around $\varphi \approx \pi/2$.

This scattering mechanism (along the *b*-direction) will mainly affect the transport along the *a*-axis. In fact, the Fermi velocities in the antinodal segments of the Fermi surface predominantly point towards the *a*-axis and therefore influencing primarily the conductivity along the *a*-axis (see equation S2).

A similar "orthogonality effect" is observable also for an anisotropic (nematic) Fermi surface as shown below.

## Anisotropic Fermi surface

Starting point is the electronic structure for an isotropic system, which is based on a tight-binding parametrization for cuprates from Ref. (*64*). We use the following hopping parameters: nearest neighbor $t_1 = 0.435$ eV, next nearest neighbor $t_2 = -0.19$ eV and next-next nearest neighbor $t_3 = 0.015$ eV.

An effective model describing the coupling between fermions and a nematic order parameter can be incorporated in a tight binding model by changing the nearest- ($\sim t_1$) and next-next nearest ($\sim t_3$) neighbor part of that parametrization in the following way

$$-t_1[\cos k_b + \cos k_a] \rightarrow -t_1[(1+\alpha)\cos k_b + (1-\alpha)\cos k_a]$$
$$-t_3[\cos 2k_b + \cos 2k_a] \rightarrow -t_3[(1+\alpha)\cos 2k_b + (1-\alpha)\cos 2k_a].$$

Figures S9A and S9B show the Fermi surfaces for $\alpha = 0$ and $\alpha = 0.02$, respectively, at a given chemical potential. The undistorted Fermi surface supports two scattering processes along $b$ which are associated with the quasi-nesting properties. Due to the higher density of states ($\sim 1/v_F$) around ($\pi$,0) we will see that the corresponding scattering is larger in this region. For $\alpha = 0.02$ (nematic state) the hole states at ($\pi$,0) are pushed above $E_F$ and a CDW scattering channel between the corresponding Fermi surface tips of adjacent Brillouin zones becomes more effective.

The scattering rate is obtained from the imaginary part of the self-energy when holes are coupled to CDW fluctuations of the standard Ginzburg-Landau form, see e.g. Ref. (*46*). One obtains

$$\Gamma_{cdw} = Im\,\Sigma(k,\omega=0) = g^2 \int \frac{d^2q}{(2\pi)^2} \frac{\varepsilon_{k-q}[f_+(\varepsilon_{k-q}) + f_-(\varepsilon_{k-q})]}{[m + \nu\eta_q - \varepsilon_{k-q}^2/\Omega]^2 + \varepsilon_{k-q}^2}$$

where $f_\pm$ denote Bose- and Fermi functions, $g$ is a coupling parameter, $m$ denotes the mass of the CDW fluctuations and $\Omega$ is a frequency cutoff. Information about the CDW momentum structure comes from $(2\pi)^2\eta_q = 4 - 2\cos(q_a - Q_a^c) - 2\cos(q_b - Q_b^c)$ with the CDW vector $\vec{Q}^c = (Q_a^c, Q_b^c)$. We take the same parameters as in Ref. (*46*). In particular, we consider here the resistivity of the 10 nm films where only CDW scattering along the $b$-direction is observed, i.e. we set $Q_c^a = 0$, $Q_c^b = 0.31 \times (2\pi)$ for a lattice constant $a \equiv 1$.

Figure S9C reports the resulting scattering rate for isotropic and anisotropic Fermi surfaces ($\alpha = 0$ and $\alpha = 0.02$). The dominating scattering rate around $\varphi \approx 0.4\pi$ between two Brillouin zones gets even more pronounced for the anisotropic Fermi surface ($\alpha = 0.02$). At the same time the Fermi velocity and the kernel entering the Boltzmann equation shown in Figure S9D around $\varphi \approx 0.4\pi$, where CDW scattering is dominant, have larger values along the $a$-direction. Hence, a unidirectional CDW along the $b$-axis will predominately affect the transport along the $a$-axis (with little to no effect on the $b$-axis transport). This is exemplified in Figure S9E, where the calculated resistivities along the in-plane $a$- and $b$-directions are shown assuming an anisotropic Fermi surface (see Fig. S9B, with $\Gamma_0 = 0.24$ meV, $\zeta = 28$). Here the unidirectional ($Q_b$) CDW scattering results in a departure from the linear-in-$T$ resistivity along the $a$-axis whereas it has no effect along the $b$-axis transport. The parameter $\beta = 0.04$ eV/K$^2$ has been chosen to fit the deviation of $\rho_a$ from linearity around $T_L$ in the 10 nm film for $p=0.12$.

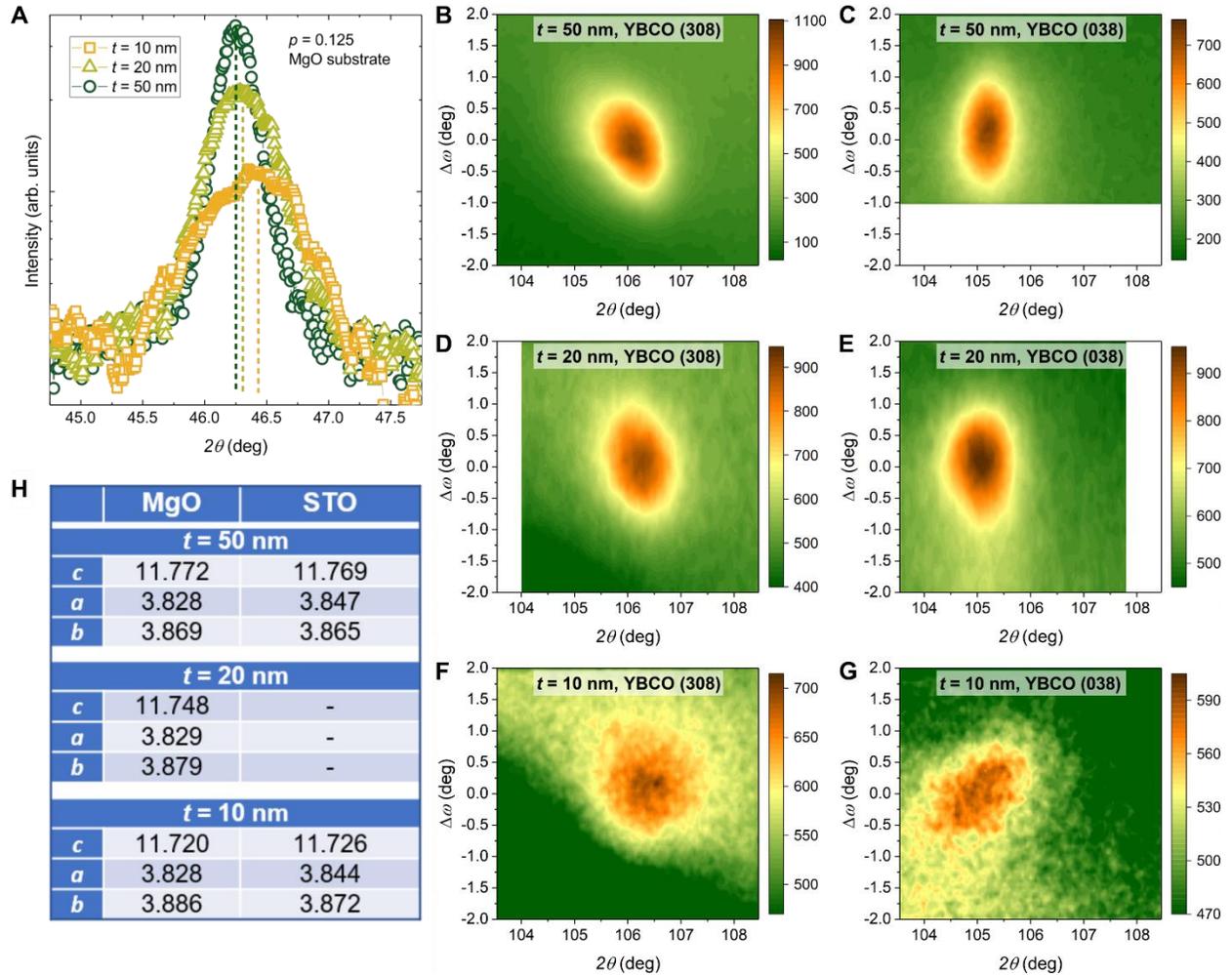

**Fig. S1**

**X-ray diffraction measurements and determination of the lattice parameters of the untwinned YBCO thin films with $p=0.125$.** **(A)** Symmetric $2\theta$-$\omega$ scans of the YBCO (006) Bragg peak for films with different thickness grown on MgO. From $2\theta_n$, maximum of the (00$n$) Bragg peak, the $c$-axis lattice parameter can be determined as $c=\lambda n/2\sin(\theta_n)$, where $\lambda=1.540598$ Å is the wavelength of the incident x-ray wave and n is the peak order. The $c$-axis presented in Fig. 1A of the main text is the average of $n=5$ and $n=6$: its shrinking, as the film thickness is reduced, is associated to the peak maximum, moving to higher $2\theta$ value. This is highlighted by the dashed lines. **(B)-(G),** Asymmetric $2\theta$-$\omega$ maps of the YBCO (308) and (038) Bragg peaks for films with different thickness grown on MgO. The $a$- and $b$-axis lattice parameters are determined from the $2\theta$ angle corresponding to the maximum respectively of the (308) and of the (038) peak (*24*). A reduction of the $c$-axis should in principle bring to an elongation of both the $a$- and $b$-axis lattice parameters, if the (308)/(038) peak positions are unchanged. However, as the film thickness is reduced, the (308) peak shifts, as the (00$n$) peaks, to higher $2\theta$ value, resulting in an $a$-axis

parameter, whose length is independent of the film thickness. The (038) peak presents instead an opposite shift, moving to lower $2\theta$ value: the $b$-axis significantly elongates, as shown in Fig. 1A. **(H)** Table with the values in Å of the lattice parameters of YBCO films grown on MgO and STO, determined from the XRD measurements.

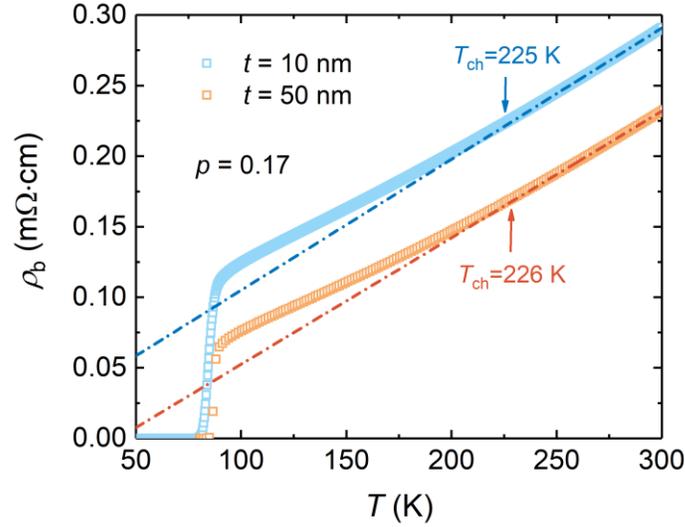

**Fig. S2**

**Influence of the CuO chains in the transport properties of thin and thick YBCO films.** $\rho(T)$ measured along the $b$-axis in $t=10$ nm and 50 nm slightly overdoped films. In the overdoped regime, along the $a$-axis, the high temperature linear resistivity is broken by an upturn, corresponding to the coherence temperature $T_{coh}$ and to a restoration of the Fermi-Liquid like behavior. Along the $b$-axis, however, at this level of doping the chains contribution to the transport, adding at any temperature a quadratic term to the resistivity, become significant (*26*). As a consequence of that, the departure from the linear-in-$T$ resistance occurs at a temperature value, $T_{ch} > T_{coh}$, which is strongly related to the CuO chains. A variation of the chains in the system, i.e. as a function of the film thickness, is expected to cause a strong variation of $T_{ch}$. In our panel, the dashed lines are the linear fits for $T>250$ K and $T_{ch}$ is the temperature where the resistivity, normalized to 290 K, deviates by 1% from the linear fit. Since $T_{ch}$ is the same in the thin and thick case, we can conclude that the CuO chains are not different in the thin and thick films and thus cannot be at the origin of the increase of in-plane anisotropy observed in the $\rho(T)$ of ultrathin underdoped films.

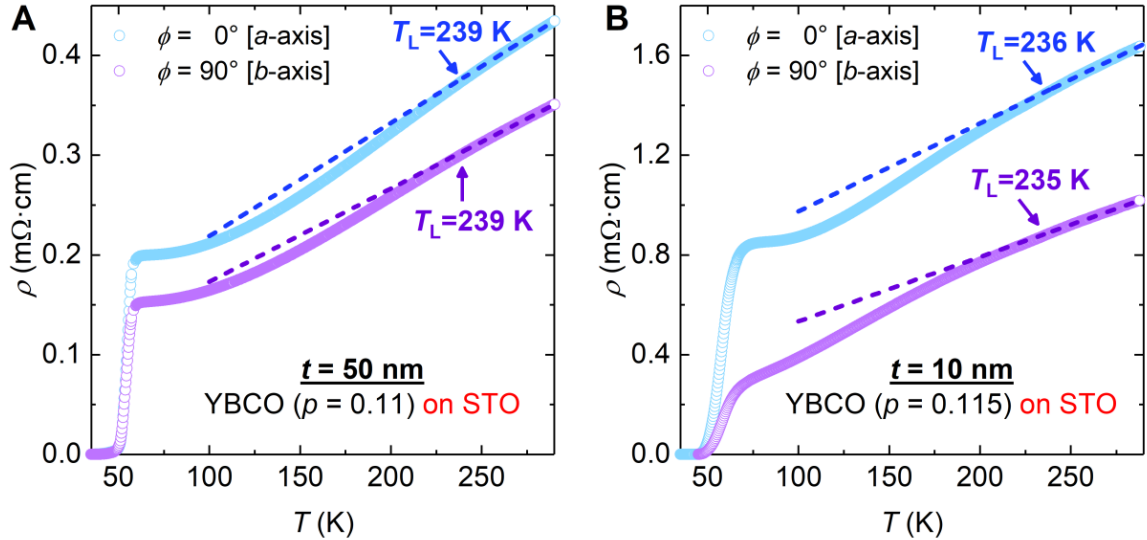

**Fig. S3**

**Angular dependence of the YBCO in-plane resistivity as a function of the thickness in films on STO.** (**A**) $\rho(T)$ of two devices, oriented along the *a*- and the *b*-axis directions, patterned on a 50 nm thick underdoped ($p$=0.11) film on STO substrate. At $T = 290$ K, $\rho_a/\rho_b = 1.2$. The dashed lines are the linear fits of the curves for $T>250$ K. $T_L$ is the temperature where the resistivity normalized to 290 K deviates by 1% from the linear fit $\rho_L(T) = \rho_0 + \gamma T$, and it is the same along both directions. Similarly to 50 nm thick films grown on MgO, the slopes of the linear region are rather similar in the two directions ($\gamma_a$=1.1 µΩcm/K and $\gamma_b$=0.9 µΩcm/K). (**B**) $\rho(T)$ of two devices, oriented along the *a*- and the *b*-axis directions, patterned on a 10 nm thick underdoped ($p$=0.115) film on STO. At $T = 290$ K, $\rho_a/\rho_b = 1.6$: as observed in Figure S1H, indeed, the orthorhombicity of the YBCO unit cell increases in the ultrathin films, even though less than on the MgO substrates. However, differently than in ultrathin films grown on MgO substrates: i) the range of linear-in-$T$ resistivity is unchanged in both directions, i.e. the extracted $T_L$ is the same; ii) the slopes of the high temperature linear resistivity are quite close in the two directions ($\gamma_a$=3.5 µΩcm/K and $\gamma_b$=2.6 µΩcm/K)

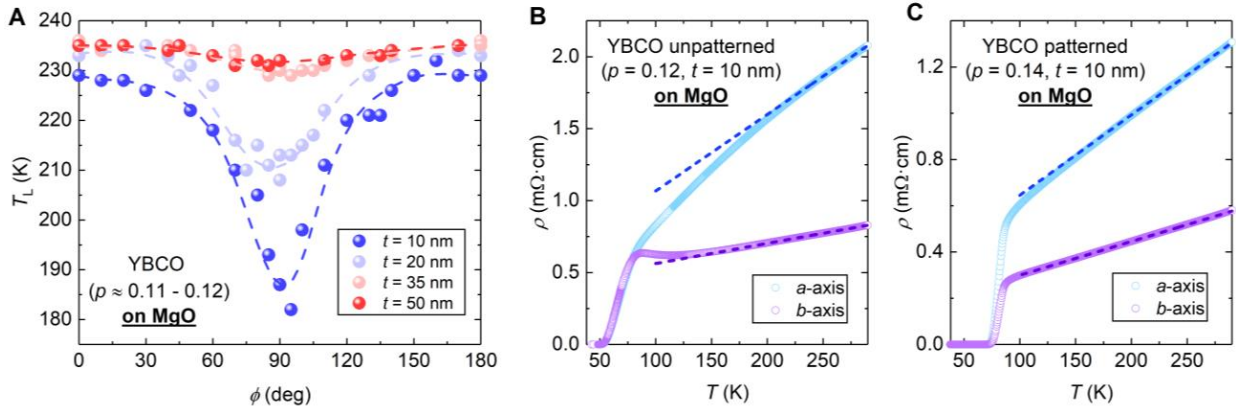

**Fig. S4**

**Angular dependence of $T_L$ in YBCO films on MgO substrates as a function of the thickness.**
(**A**) The $T_L$, extracted as described in Figure 2, is presented as a function of $\phi$ for four films ($p \approx 0.11$-$0.12$) with thickness $t$ ranging from 10 to 50 nm. As the thickness is reduced, the value of $T_L$ measured along the $b$-axis ($\phi = 90°$) gradually decreases. This implies that the linearity of $\rho$ mainly extends to the whole normal state of YBCO, when the thickness is reduced. Moreover, we observe that $T_L$ is approximately constant from $\phi = 0°$ up to about $\phi = 45°$, followed by a sharp suppression with a minimum at $\phi = 90°$. From this occurrence we infer that the scattering mechanism – if any - responsible of the extension of the linear regime to lower $T$ must be strongly directional. In this figure, the doping level of the $t = 10$ nm film is slightly higher, resulting in a lower value of $T_L$ at $\phi = 0°$. The dashed lines are guides to the eye. (**B**) $\rho(T)$ measured along the $a$- and the $b$-axis directions of a 10 nm thick underdoped ($p = 0.12$) film. With respect to the $\rho(T)$ presented in Figure 2B, here the film has not been patterned into microwires, therefore the current paths are not well defined and a proper angular dependence of the resistivity cannot be measured. However, both the striking features we have observed in the $\rho(T)$ of our ultrathin films on MgO, i.e. the enhanced in-plane anisotropy ratio of the resistivity $\rho_a/\rho_b$ at 290 K and the much broader temperature range of linearity along the $b$-axis, are present. This occurrence confirms that the results we can infer from the $\rho(T)$ are robust and do not depend on possible damages induced into the ultrathin film by the nanopatterning procedure. (**C**) $\rho(T)$ of two devices, oriented along the $a$- and the $b$-axis directions, patterned on a 10 nm thick underdoped ($p = 0.14$) film. Along the $b$-axis, $T_L = 101$ K, i.e. close to the onset of superconducting fluctuations.

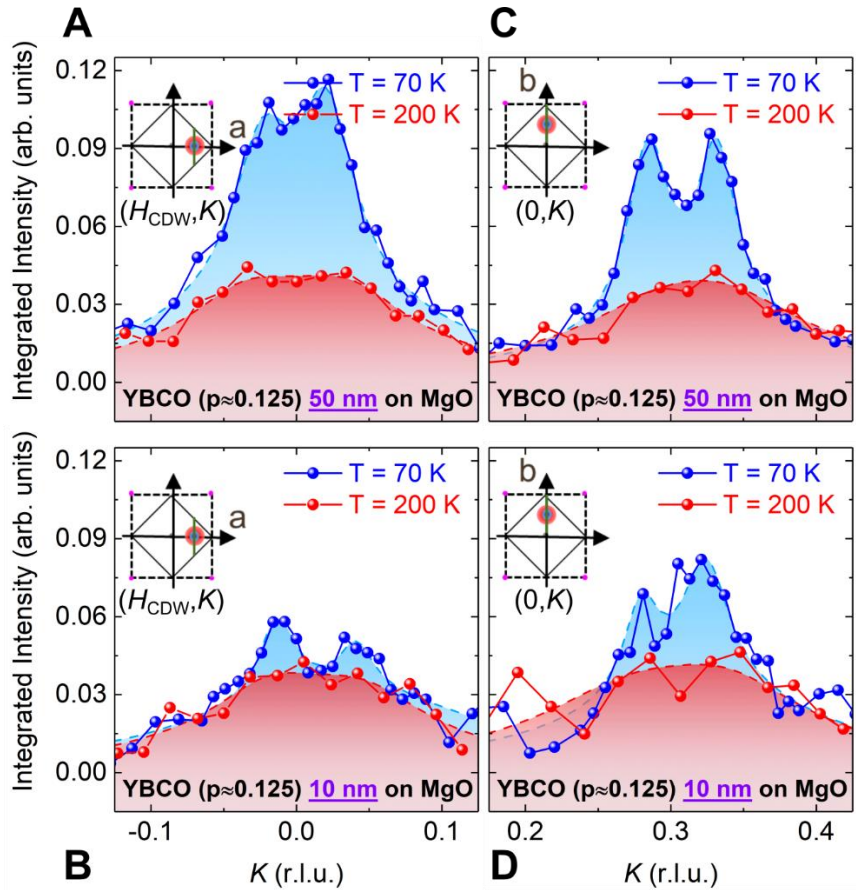

**Fig. S5**

**Thickness dependence of the CDW in YBCO thin films, K scans.** The same scans presented in Figure 4 along the *H* direction of the reciprocal space are here shown along the *K* direction. **(A), (C)** Quasi-elastic scans measured at $T = 70$ K and $T = 200$ K on the 50 nm thin film on MgO along the ($H_{CDW}$,K) direction, i.e. *a*-axis (A), and the (0,K) direction, i.e. *b*-axis (C). **(B),(D)** Same as panels (A) and (C), but on a 10 nm thick sample. In the latter sample, the CDW intensity along the *a*-axis is almost negligible, similarly to what observed along the *H* direction. The CDW peaks are split in the reciprocal space: this is a consequence of the buckling of the atomic planes along the *b*-direction, occurring in YBCO films as a consequence of the large lattice mismatch with the MgO substrates (*24*). On each film, we notice that the maximum intensity of the CDW peak along the *a*-axis is the same when measured along the *H* and the *K* direction (when split, $H_{CDW}$ and $K_{CDW}$ are chosen to be in the middle between the two maxima). Same applies for the *b*-axis. This occurrence confirms that the drop of intensity we have observed along the *a*-axis in ultrathin films on MgO cannot be due to a misalignment, but it is an intrinsic effect of the investigated samples.

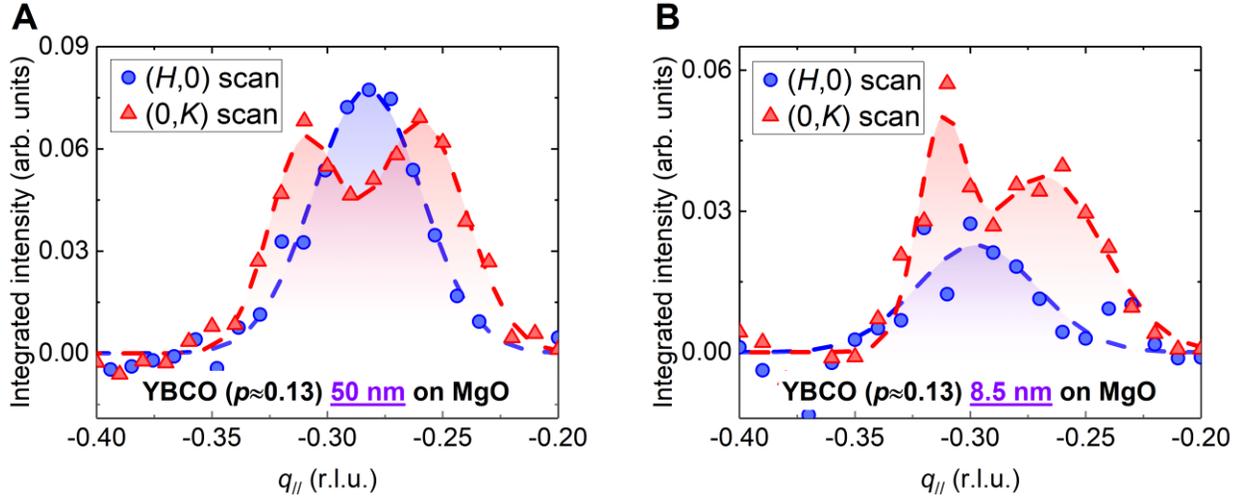

**Fig. S6**

**Doping evolution of the CDW vs thickness in YBCO films on MgO.** RIXS spectra have been measured on two underdoped YBCO films, with thickness of 50 nm and 8.5 nm, grown on MgO substrates. With respect to those presented in Figure 4 and Fig. S5, these films are characterized by a higher oxygen doping of $p \approx 0.13$ ($T_c \approx 75$ K). **(A)** Quasi-elastic scans measured at $T = 20$ K on the 50 nm thin film on MgO both along the ($H$,0) direction, i.e. $a$-axis (circles) and along the (0,$K$) direction, i.e. $b$-axis (triangles). Here the splitting of the CDW peaks along the (0,$K$) direction is related to the unidirectional buckling of the $CuO_2$ planes along the $b$-axis, previously discussed. **(B)** Same as panel (A), but on a 8.5 nm thick sample. In the latter sample, the ratio $CDW_{(H,0)}/CDW_{(0,K)}$ is lower than in the thicker film as a consequence of the reduction of the charge order along the $a$-axis. This is in agreement to what already observed on the $p \approx 0.125$ sample in the main text. With respect to that sample, here the reduction is not as strong.

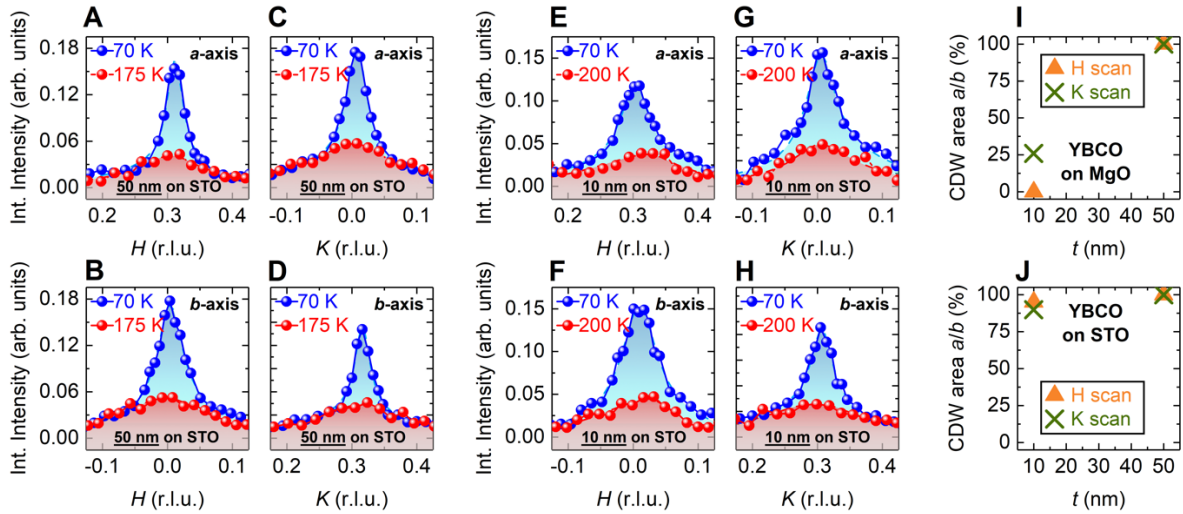

**Fig. S7**

**CDW as a function of thickness in YBCO on STO and comparison with films grown on MgO.**
**(A),(B),(C),(D)** Quasi-elastic scans measured at $T = 70$ K and $T = 175$ K on the 50 nm thin film ($p \approx 0.125$) on STO along the *a*-axis [($H$,0) and ($H_{CDW}$,$K$) directions, (A)-(C)] and *b*-axis [($H, K_{CDW}$) (0,$K$) directions, (B)-(D)]. Each scan is shown after subtracting the background, given by the quasi-elastic scan measured along the diagonal [($H$,$H$) direction]. **(E),(F),(G),(H)** Same as the first four panels, but on a 10 nm thick film ($p \approx 0.125$) on STO substrates. In the latter sample, differently from the film grown on MgO, the CDW intensities along the *a*-axis and *b*-axis are comparable. **(I),(J)** The CDW area, determined as the difference between the low temperature and the high temperature quasi-elastic scans measured along the $H$ (triangles) and $K$ (crosses) directions, and normalized to the *b*-axis contribution, is plotted as a function of the thickness for films both grown on MgO (I) and STO (J). The values have been obtained after removing for each direction the signal coming from the twin domains ($\approx$15% for films grown on MgO, $\approx$10% for films grown on STO). The RIXS measurements tells that on MgO substrates, the CDW signal along YBCO *a*-axis is suppressed in the 10 nm thick films. Here the CDW is unidirectional along the *b*-axis. On STO substrates, on the contrary, the CDW signal along the YBCO *a*-axis is approximately equal to that along the *b*-axis, both on the 10 nm and 50 nm thick films, as expected for YBCO at that level of doping.

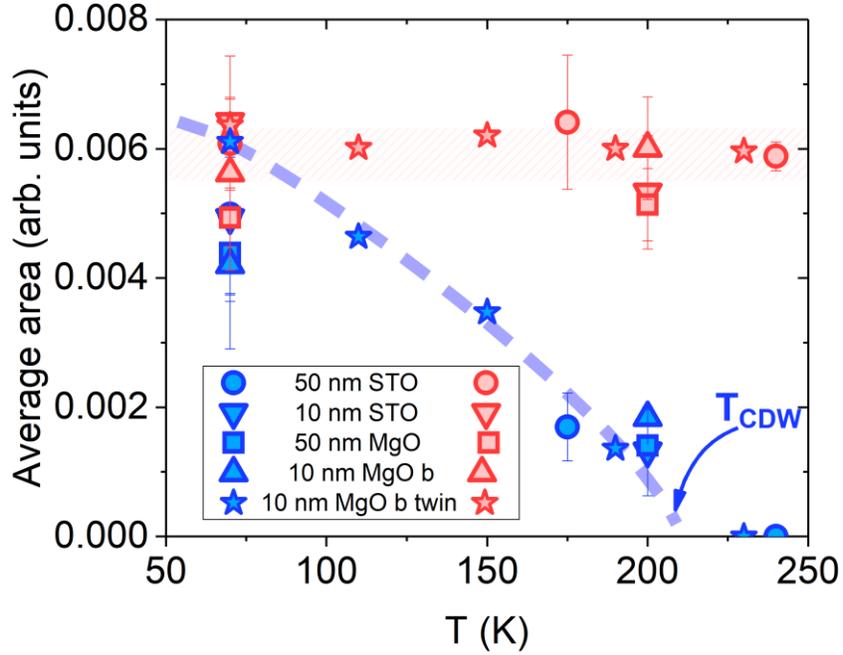

**Fig. S8**
**Temperature dependence of the CDW in YBCO films with $p=0.125$.** The quasi-elastic scans, measured on the four samples presented in Figures 4, S5 and S7, and on a fifth sample, 10 nm thick YBCO on MgO, twin of that presented in Figures 4C, 4D, S5B, S5D, have been fitted using an almost $T$-independent Lorentzian associated to CDF (*43*) and a $T$-dependent Lorentzian associated to CDW. Finally the areas of all these peaks have been plotted in the plot above (blue symbols for CDW, red symbols for CDF) as a function of the temperature. Despite the different used samples, film thicknesses and substrates, all the different data sets are consistent not only qualitatively but also quantitatively, and they show a similar trend. In particular the areas of all the CDF peaks lay in a narrow range, and the variance of their distribution, $\Delta$ (given by the height of the shaded red area), can be considered a measure of the uncertainty of our measurements. The CDW peak is $T$-dependent and at 190 K is still present, being characterized by an area, which is higher than our uncertainty $\Delta$. From a second order polynomial fit of the CDW area vs $T$ of the 10 nm thick film on MgO (blue stars) we have determined $T_{CDW} \approx 208$ K, which is very close to the $T_L$ we have measured at this level of doping, and close to the $T^*$ measured by other techniques at this level of doping (*8,27*). Note that the areas of the CDW peaks measured along the *a*-axis directions on the 10 nm thick films on MgO, not included in the plot above, are at all the temperatures above $T_c$ smaller than $\Delta$ (after considering the contribution of twinning domains). The CDW in these ultrathin films is therefore unidirectional and oriented along the *b*-axis direction.

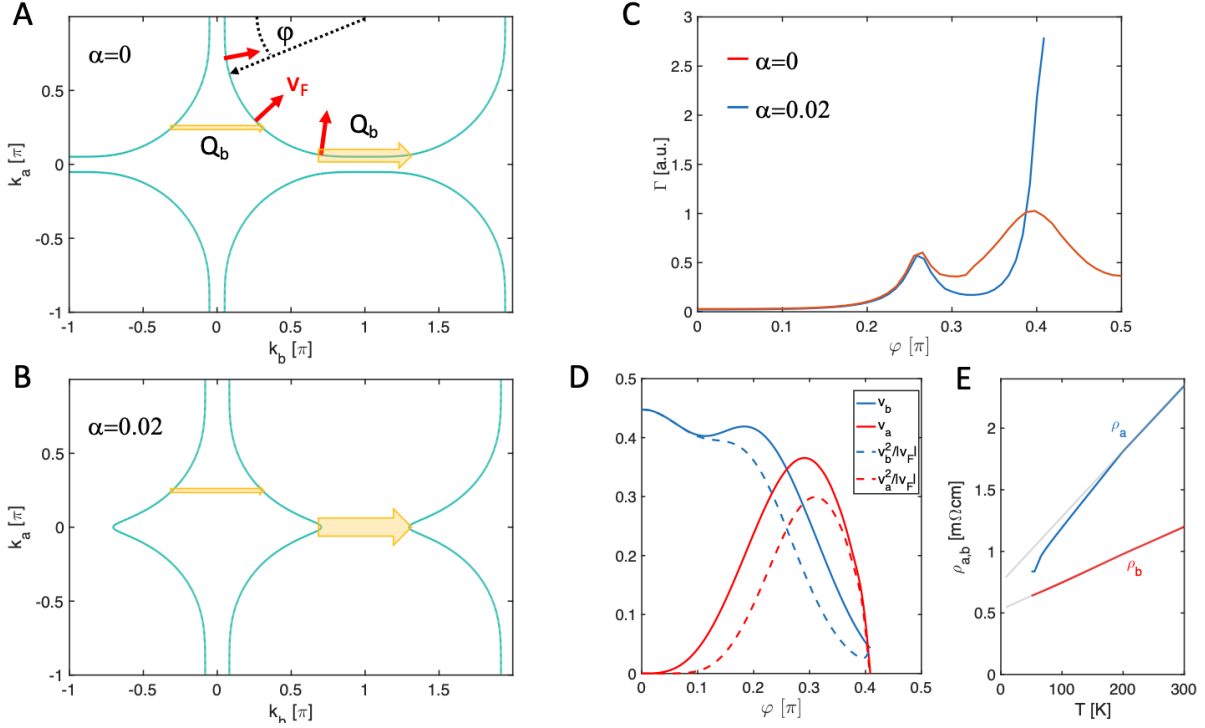

**Fig. S9**

**Effect of a distorted Fermi surface on the CDW and on the transport.** Fermi surfaces for α=0 (**A**) and α=0.02 (**B**). The dominating CDW scattering processes are shown as orange arrows. The red arrows represent the Fermi velocity vectors pointing orthogonal to the Fermi surface. (**C**) Angular dependence of the CDW scattering rate for $\alpha = 0$ (red) and $\alpha = 0.02$ (blue), see panel (A) for the definition of $\varphi$. (**D**) Fermi velocity and kernel entering the Boltzmann equations along the *a*- and *b*- direction. (**E**) Resistivity calculated from the Boltzmann equations along the *a*- and *b*-direction. The grey lines are linear extrapolations of the high temperature resistivity.